\documentclass[sigconf,language=english]{acmart}

\AtBeginDocument{%
  }

\setcopyright{acmlicensed}
\copyrightyear{}
\acmYear{}
\acmDOI{XXXXXXX.XXXXXXX}
\acmISBN{978-1-4503-XXXX-X/2018/06}
  
\acmSubmissionID{2026-12*}
\usepackage{algorithm}    
\usepackage{algpseudocode}
\usepackage{amsmath}      
\usepackage{float}
\usepackage{cleveref}
\usepackage{multirow}
\usepackage{tabularx}
\usepackage[table]{xcolor} 
\usepackage{graphicx}
\usepackage{subcaption}
\usepackage{xcolor}
\usepackage{pifont} 

\begin{document}

\title{ASARL: Autonomous Social-Aware Relevance Learning for QQ Search}


\author{Tao Su}
\authornotemark[1]
\email{victorsu@tencent.com}
\affiliation{%
  \institution{Tencent PCG}
  \city{Beijing}
  \country{China}
}

\author{Jinjing Hu}
\authornote{Equally contribution.}
\authornote{Corresponding author.}
\email{jinjinghu@tencent.com}
\affiliation{%
  \institution{Tencent PCG}
  \city{Beijing}
  \country{China}
}

\author{Xiao Wang}
\email{leonsiuwang@tencent.com}
\affiliation{%
  \institution{Tencent PCG}
  \city{Beijing}
  \country{China}
}

\author{Xingzhong Cao}
\email{maxwellcao@tencent.com}
\affiliation{%
  \institution{Tencent PCG}
  \city{Beijing}
  \country{China}
}

\author{Hui Wang}
\email{joltwang@tencent.com}
\affiliation{%
  \institution{Tencent PCG}
  \city{Beijing}
  \country{China}
}


\begin{abstract}
The rapid growth of online social platforms has transformed communication and information retrieval, giving rise to social search, where queries-titles are typically expressed in informal, community-specific language. While large language models provide strong general-purpose semantic understanding, their effectiveness in social search is constrained by contextual discrepancy, data scarcity, and behavior-driven dynamics. To address these challenges, we propose the \textbf{Autonomous Social-Aware Relevance Learning (ASARL)}, a fully automated framework that integrates multi-agent data curation with staged model training. ASARL leverages a collaborative agent system: \textit{ReasonAgent} generates interpretable relevance labels grounded in social attributes, \textit{CriticAgent} validates and ensures logical consistency, and \textit{GenAgent} augments long-tail data through synthetic query–title pairs. Building on the curated dataset, ASARL employs three-stage training: \textit{Social Context Training (SCT)} to capture social language patterns, \textit{Preference-Guided Optimization (PGO)} to align model predictions with behavioral signals, and \textit{Social Distillation (SD)} to transfer these improvements into compact models for efficient deployment. Extensive offline and online experiments on the QQ search platform demonstrate significant improvements in both offline relevance metrics and online user engagement indicators, along with enhanced annotation efficiency. These results validate the effectiveness of combining autonomous, socially grounded data governance with preference-aligned training in practical search systems.
\end{abstract}


\ccsdesc[500]{Information systems~Relevance assessment}
\ccsdesc[500]{Information systems~Language Models}
\ccsdesc[500]{Computing methodologies~Natural language processing}

\keywords{Search Relevance, Multi-Agent Co-operation, Large Language Model}


\maketitle


\begin{figure}[t]
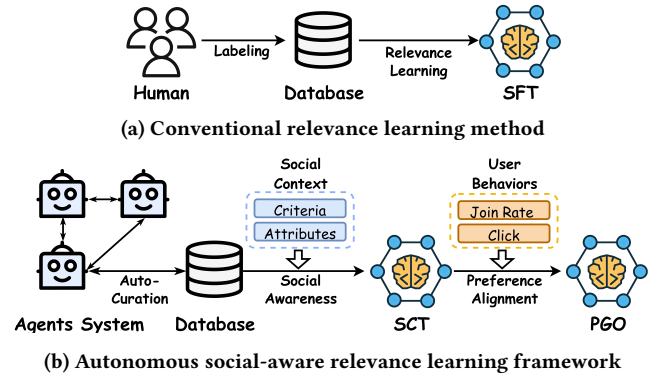

  \centering
  \begin{subfigure}[b]{\columnwidth}
    \centering
    \includegraphics[width=\columnwidth]{Img/a.png}
    \caption{Conventional relevance learning method}
    \label{fig:system_a}
  \end{subfigure}

  \vspace{0.5em}

  \begin{subfigure}[b]{\columnwidth}
    \centering
    \includegraphics[width=\columnwidth]{Img/b.png}
    \caption{Autonomous social-aware relevance learning framework}
    \label{fig:system_b}
  \end{subfigure}

  \caption{Comparison between (a) a conventional relevance learning method, which relies on manual labeling and direct model training, and (b) our autonomous social-aware relevance learning framework, which integrates automated data governance, social-aware modeling, and preference alignment.}
  \label{fig:system_comparison}
\end{figure}

\section{Introduction}
The widespread adoption of online social platforms has reshaped communication and information interaction, leading to the emergence of \textit{social search}. Platforms such as QQ, Twitter, and Instagram host diverse interest groups, communities, and user generated content, which differ fundamentally from conventional web\cite{liu2021pre,li2025towards,brin1998anatomy} and e-commerce search\cite{chen2025onesearch,peng2024large,Yu2024}. While web search primarily targets general purpose information retrieval and e-commerce search focuses on transactional queries, social search centers on community driven content and interest-based interactions. Queries in this context are often informal, containing slang, colloquial expressions, or subcultural references, which deviate from standard query formats and pose challenges for conventional search systems.

Traditional search approaches, including embedding-based ranking methods\cite{huang2020embedding,wu2024llm} and Transformer-based models\cite{devlin2019bert,he2020deberta,liu2019roberta,yao2022reprbert}, rely on abundant labeled data and well-structured query representations. These methods face three key limitations in social search scenarios: contextual discrepancy., as informal and community-specific language hinders alignment with standard representations; data scarcity, particularly for long-tail or niche interest groups; and behavior-driven dynamics, where user interactions provide feedback that static annotations cannot capture. Consequently, conventional models often produce suboptimal relevance predictions and fail to cover diverse social queries effectively.

Recent advances in large language models (LLMs) have improved relevance modeling in industrial search, enabling direct relevance comparison \cite{fang2025adore,tang2025lref}, structured information extraction \cite{tian2025towards,zhao2025explainable}, and knowledge distillation \cite{shang2025knowledge,li2025proactive,wu2025large}. Yet these methods, designed for transactional or web domains, typically rely on a single model for data generation or prediction, limiting adaptability in dynamic social search environments. This motivates the need for frameworks capable of scalable annotation, structured reasoning, and alignment with behavioral feedback, motivating subsequent developments in automated and socially aware search systems.

To address these challenges, we propose the \textbf{Autonomous Social-Aware Relevance Learning (ASARL)}, a fully automated framework that integrates multi-agent data curation with staged alignment training. ASARL employs three specialized agents: \textbf{ReasonAgent} generates fine-grained relevance labels and reasoning traces for query–title pairs; \textbf{CriticAgent} validates the annotations and ensures social and logical consistency; and \textbf{GenAgent} autonomously generates candidate titles for underrepresented regions, mitigating long-tail scarcity. Through closed-loop collaboration, these agents produce a high-quality, socially representative dataset, which forms the foundation for subsequent model training.

Building on this curated dataset, ASARL adopts a three-stage training pipeline. First, \textbf{Social Context Training (SCT)} equips the model with core social intents and attributes awareness derived from curated annotations. Then, \textbf{Preference-Guided Optimization (PGO)} further aligns the model with real user behaviors, leveraging signals such as click-through and join rates. Finally, \textbf{Social Distillation (SD)} transfers these capabilities into compact student models, enabling efficient and scalable deployment.

As shown in \Cref{fig:system_comparison}, conventional relevance learning methods rely on manual labeling and single-stage training, which limits scalability and model adaptation to social search scenarios. In contrast, ASARL autonomously integrates agent-based data curation, social-aware model foundation training, and preference-aligned optimization. This automated framework ensures high-quality, socially grounded annotations and effectively incorporates user feedback, moving beyond simple semantic matching while maintaining interpretability and social relevance in search results.

Through extensive experiments, we show that ASARL consistently improves relevance prediction in social search. On offline benchmarks, it achieves higher Accuracy, Macro F1, and NDCG compared with baseline models, demonstrating better alignment with ground-truth relevance labels. Online evaluations further confirm the practical benefits of our approach: ASARL increases Click-Through Rate, Join Rate, and GSB, indicating more engaging and socially aligned search results. Moreover, the multi-agent system substantially improves annotation efficiency, particularly for long-tail topics, validating the feasibility of fully autonomous, social-aware data curation and model training.

We summarize the key contributions of our work as follows:
\begin{itemize}
    \item \textbf{Problem Insight}: We identify key challenges of social search and the limitations of relevance learning methods.  
    \item \textbf{Methodological Innovation}: We propose ASARL, an autonomous social-aware relevance learning framework that generates interpretable data, inject social context, and aligns models with user preferences.  
    \item \textbf{Practical Validation}: We demonstrate on the QQ search platform that ASARL improves relevance metrics and annotation efficiency.  
\end{itemize}

\section{Related Work}

\subsection{Traditional Search Relevance Methods}
Transformer-based models, such as BERT\cite{devlin2019bert} and its variants\cite{he2020deberta, liu2019roberta, yao2022reprbert, santhanam2021colbertv2}, have become the backbone of traditional relevance modeling in both web search\cite{liu2021pre,li2025towards,brin1998anatomy} and e-commerce\cite{chen2025onesearch,peng2024large,Yu2024}. These models are typically fine-tuned on large-scale query–document pairs, where human annotation links user queries to relevant documents such as product descriptions or articles. By capturing both lexical and contextual alignment, Transformers have significantly advanced ranking performance, leveraging structured datasets and abundant user interaction signals\cite{ai2017learning,guo2019attentive}.

Despite their success, these methods face critical limitations in complex search scenarios. Manually annotated data is costly and sparse, particularly for long-tail queries\cite{peng2024large,zhu2024collaborative} and domain-specific topics. Moreover, the supervised framework underlying these models struggles with the informal, fragmented, and rapidly evolving nature of online content. As a result, traditional Transformer-based approaches are often unable to generalize effectively to the dynamic and data-scarce environment of social search. These challenges have motivated the exploration of large language models (LLMs), which offer stronger generative capabilities and greater adaptability.

\subsection{LLMs for Search Relevance}
The rise of large language models\cite{yang2025qwen3,liu2024deepseek,agarwal2025gpt} has opened new possibilities for addressing the bottlenecks of data annotation and model training. LLMs can automatically generate labeled datasets through techniques such as self-generated data and chain-of-thought (CoT) reasoning\cite{wei2022chain,zhang2022automatic,li2024chain}, thereby reducing dependence on manual annotation while improving data quality. They also play a central role in training pipelines: supervised fine-tuning\cite{Alpaca2023,hsieh2023distilling}, reinforcement learning from preferences\cite{dpo,kto,grpo}, and knowledge distillation\cite{shang2025knowledge,li2025proactive,wu2025large} have all been used to enhance generalization and efficiency, enabling deployment in resource-constrained environments.

Recent studies have applied these capabilities to search systems. LLMs have been employed to enrich training corpora by generating synthetic queries and hard samples\cite{li2024syneg}, to refine ranking with CoT-driven reasoning\cite{fang2025adore,tang2025lref}, and to distill knowledge into smaller relevance models for real-time serving\cite{shang2025knowledge,li2025proactive,wu2025large}. They also contribute to explainability by transforming complex reasoning into interpretable outputs\cite{tian2025towards,zhao2025explainable}, and have improved multimodal search tasks such as video retrieval through integration with speech recognition\cite{zhang2025efficient}. Personalized search has further benefited from LLMs that align recommendations with user intent and motivation\cite{qin2025maps}.  

While promising, most of these approaches rely on a single LLM either to act as a relevance model or to generate auxiliary training data. Such single-agent paradigms limit adaptability and robustness, as they cannot exploit collaborative mechanisms or iterative feedback. This motivates the development of multi-agent systems that orchestrate different roles to improve both data quality and model performance.  

\begin{figure}[t]
  \centering
  \includegraphics[width=\linewidth]{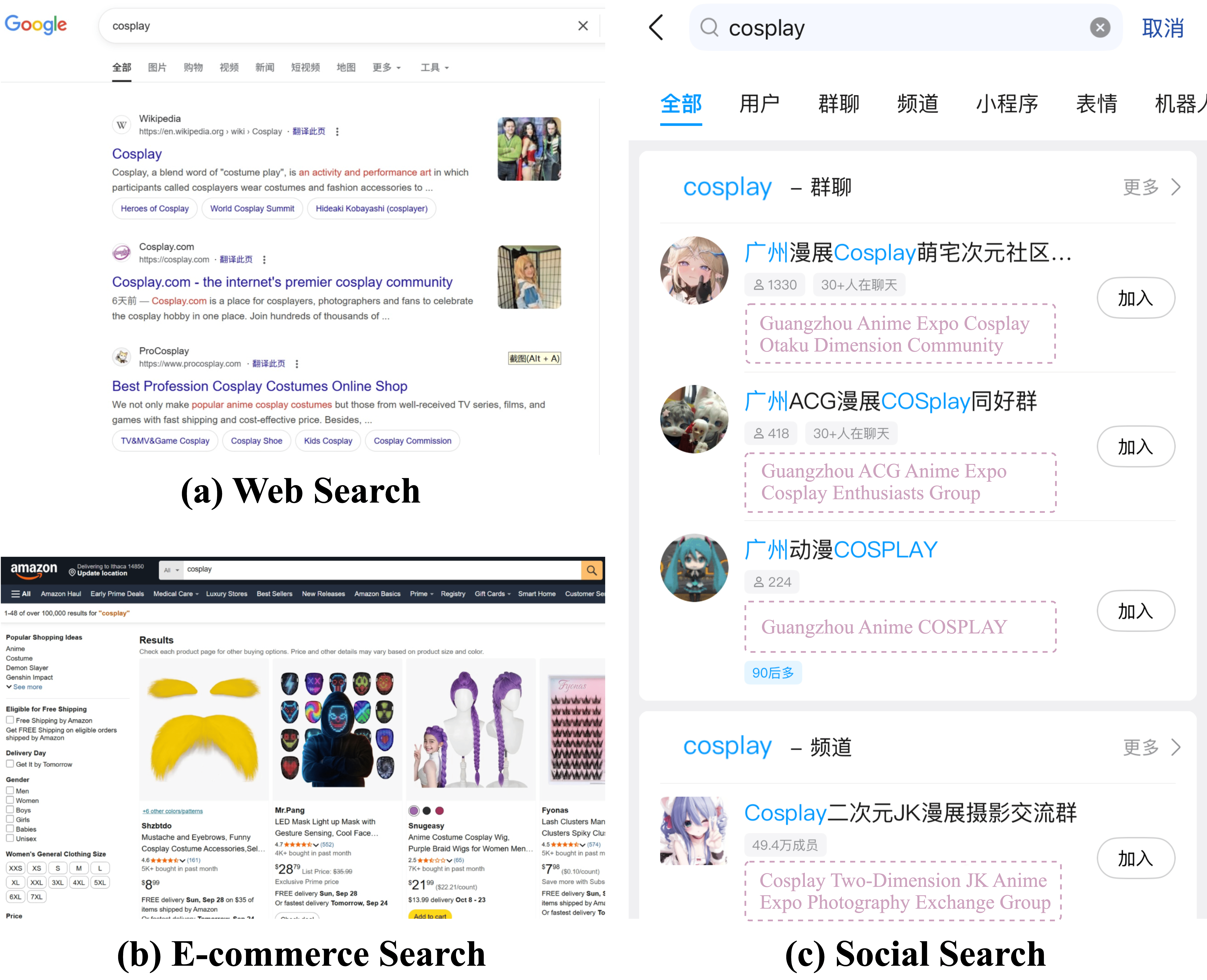}
  \caption{Illustration of different search paradigms: (a) conventional web search, (b) e-commerce search, and (c) social search on the QQ platform, which centers on user-generated groups and channels.}
  \label{fig:social_search}
\end{figure}

\subsection{Multi-Agent Systems for Search Relevance}
Multi-agent systems (MAS)\cite{li2024survey} provide a natural framework for collaborative problem-solving, where multiple agents interact to perform tasks such as data curation\cite{li2024autodcworkflow,qi2024cleanagent}, planning\cite{yang2025agentnet,li2024agent}, and model optimization\cite{hong2024metagpt,li2023camel,yuan2024evoagent}. By leveraging the complementary strengths of individual agents, MAS have been successfully applied to domains requiring large-scale automation and decision-making. In recent work, LLM-powered agents have been used for data cleaning\cite{li2024autodcworkflow,qi2024cleanagent}, strategy generation\cite{motwani2024malt,mallampati2025dynamicstrategy}, and knowledge distillation\cite{kang2025agentdistillation,chen2024magdi}, demonstrating their potential for scalable workflows.

However, applications of MAS to search relevance remain limited. Existing practices typically use a single LLM to generate labeled data or directly serve as a ranking model, overlooking the advantages of agent collaboration. To fill this gap, we propose a multi-agent framework tailored for relevance search. Our system coordinates three specialized agents—GenAgent, CriticAgent, and ReasonAgent—that iteratively curate training data and refine model outputs. The curated data is then used to train a relevance model, which undergoes a three-stage alignment process for further optimization. This feedback-driven, self-sustaining pipeline ensures continuous improvement in both data quality and model robustness, addressing the persistent challenges of data scarcity and evolving user preferences in dynamic search environments.

\section{Preliminaries}
In this section, we formally introduce the task of social search and highlight its distinct characteristics compared with conventional web or e-commerce search. Specifically, given a user query $q$ and a candidate group or channel title $t$, the goal is to predict the relevance score that reflects how well the candidate matches the query intent.  

Unlike traditional search, social search involves user-generated communities and interest channels, which bring unique challenges (see \Cref{fig:social_search}):  

\begin{itemize}
    \item \textbf{Contextual Discrepancy.} Social search titles often feature informal language or community-specific terms, deviating from standard query formats and involving complex social context, making them difficult to align with traditional methods.
    \item \textbf{Data Scarcity.} Many long-tail communities lack sufficient annotated training data, which limits the generalization ability of supervised relevance models.
    \item \textbf{Behavior-Driven Dynamics.} Relevance in social search is often better reflected by user interaction signals, which provide dynamic feedback beyond static annotated labels and motivate preference-aligned optimization.
\end{itemize}

These challenges make social search different from traditional paradigms and motivate adaptive frameworks that can handle contextual discrepancy, compensate for limited supervision, and take behavioral feedback into account to better align models with real-world usage.

\section{Method}
\begin{figure*}[t]
  \centering
  \includegraphics[width=.95\linewidth]{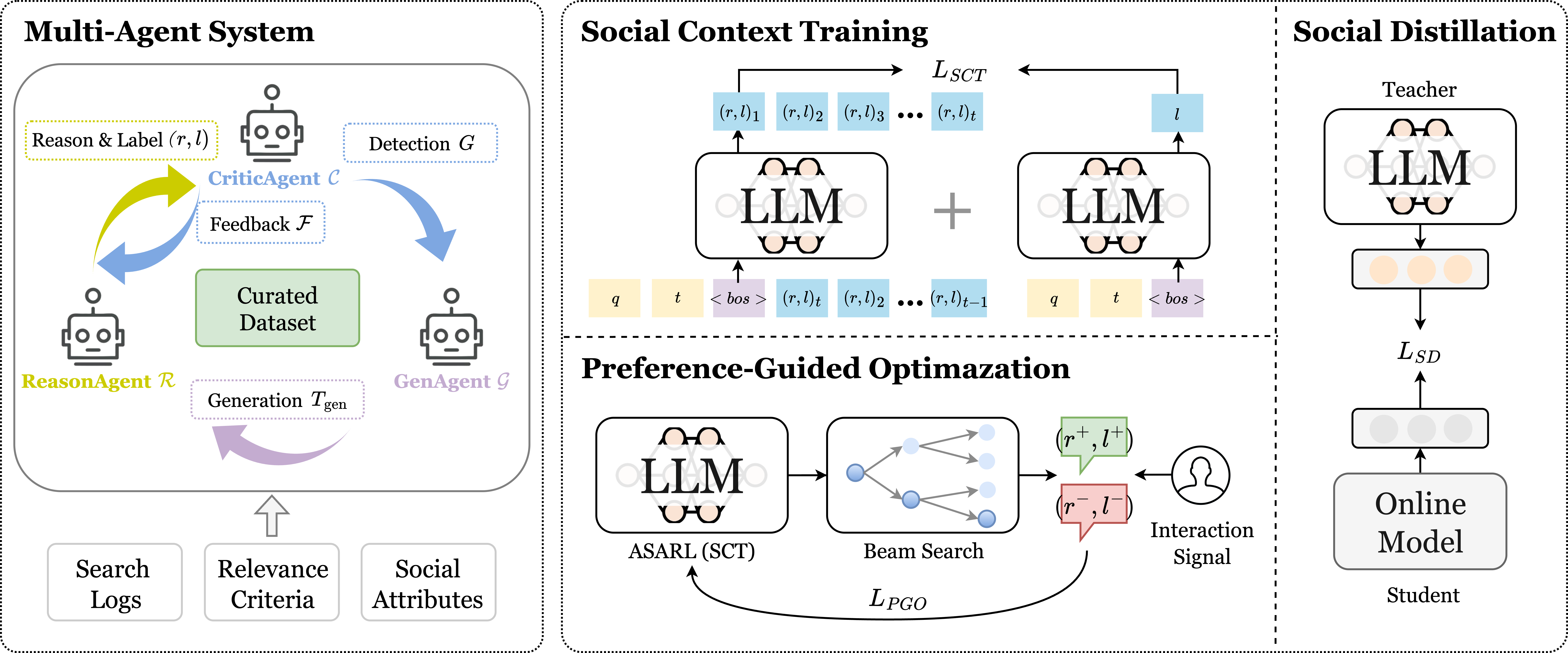}
  \caption{Overview of ASARL. Queries are annotated and validated by a multi-agent system (ReasonAgent, CriticAgent, GenAgent) in a closed feedback loop to generate a high-quality, diverse, and socially aligned dataset. The curated dataset is then used in staged alignment training (SCT, PGO, SD) to produce robust social-aware search models.}
  \label{fig:pipeline}
\end{figure*}

To address the challenges of contextual discrepancy, data scarcity, and behavior-driven dynamics in social search, we propose the \textbf{Autonomous Social-Aware Relevance Learning (ASARL)}. As shown in \Cref{fig:pipeline}, ASARL integrates \textit{multi-agent collaborative data curation} with a \textit{three-stage alignment training process}, forming a fully automated pipeline from raw data generation to socially aligned model optimization. The multi-agent system ensures that the curated dataset is both high-quality and socially representative, while the training process progressively enhances the search model's interpretability, preference alignment, and deployability.

\begin{algorithm}[t]
\caption{Multi-Agent Collaborative Data Curation in ASARL}
\label{alg:agent_collab}
\begin{algorithmic}[1]
\Require Raw query set $Q$, sampled candidates $T(q)$ for each $q \in Q$
\Ensure Curated dataset $D^*$

\State $D_{\text{init}} \leftarrow \emptyset$
\ForAll{$q \in Q$}
\State Sample $t^+$ (top), $t^\sim$ (token-based), $t^-$ (random) from $T(q)$
\ForAll{$t \in {t^+, t^\sim, t^-}$}
\State $(r, l) \leftarrow \mathcal{R}(q, t)$ \Comment{ReasonAgent: intent–attribute matching}
\State $D_{\text{init}} \leftarrow D_{\text{init}} \cup {(q, t, r, l)}$
\EndFor
\EndFor

\State $(D_v, D_{\neg v}, \mathcal{F}) \leftarrow \mathcal{C}(D_{\text{init}})$
\State $iter \leftarrow 0$ \Comment{Initialize iteration counter}
\While{$D_{\neg v} \neq \emptyset$ and $iter < maxIter$}
\State $D_r \leftarrow \mathcal{R}(D_{\neg v}; \mathcal{F})$ \Comment{ReasonAgent re-annotates using feedback}
\State $(D_v', D_{\neg v}', \mathcal{F}') \leftarrow \mathcal{C}(D_r)$ \Comment{CriticAgent re-validates}
\State $D_v \leftarrow D_v \cup D_v'$ \Comment{Accumulate validated samples}
\State $D_{\neg v} \leftarrow D_{\neg v}'$ \Comment{Update remaining invalid samples}
\State $\mathcal{F} \leftarrow \mathcal{F}'$ \Comment{Update feedback for next iteration}
\State $iter \leftarrow iter + 1$ \Comment{Increment iteration counter}
\EndWhile
\State $D_b \leftarrow D_v$ \Comment{Base dataset after Reason–Critic loop}

\State $G \leftarrow \mathcal{C}{\text{dist}}(D_b)$ \Comment{Detect underrepresented regions}
\State $D_s \leftarrow \emptyset$
\ForAll{$q \in G$}
\State $T{\text{gen}} \leftarrow \mathcal{G}(q)$ \Comment{GenAgent generates new titles}
\ForAll{$t_g \in T_{\text{gen}}$}
\State $(r_g, l_g) \leftarrow \mathcal{R}(q, t_g)$ \Comment{ReasonAgent annotates generated titles}
\If{$\mathcal{C}_{\text{check}}(q, t_g, r_g, l_g) = 1$} \Comment{CriticAgent validates}
\State $D_s \leftarrow D_s \cup {(q, t_g, r_g, l_g)}$
\EndIf
\EndFor
\EndFor

\State $D^* \leftarrow \mathcal{C}_{\text{final}}(D_b \cup D_s)$
\Return $D^*$
\end{algorithmic}
\end{algorithm}

\subsection{Multi-Agent System}
ASARL employs a collaborative multi-agent system consisting of three specialized agents: \textbf{ReasonAgent}, \textbf{CriticAgent}, and \textbf{GenAgent}, as summarized in Algorithm~\ref{alg:agent_collab}. Their complementary roles ensure that the final dataset is not only valid and diverse but also socially aware.

\subsubsection{ReasonAgent}
The ReasonAgent $\mathcal{R}$ provides initial annotations for query–title pairs by sampling candidates from search logs in three tiers: top-ranked cases $t^+$, token-matched cases $t^\sim$, and randomly selected cases $t^-$. For each pair $(q_i,t_i)$, the agent performs intent–attribute matching and outputs a relevance label $l_i \in$ \{"Relevant", "Partially Relevant", "Irrelevant"\} along with a reasoning trace $r_i$:
\begin{equation}
    (r_i,l_i) \leftarrow \mathcal{R}(q_i, t_i).
\end{equation}

Specifically, $r_i$ is a structured trace decomposing the query–title pair into predefined social intents and attributes, and $l_i$ reflects the relevance determined by this alignment. This yields annotations that are interpretable and consistent with social relevance criteria.

\subsubsection{CriticAgent}
The CriticAgent $\mathcal{C}$ serves as both a validator and a distributional monitor. Given the initial dataset $D_{\text{init}}$ from the ReasonAgent, it checks the consistency of relevance labels, reasoning traces, and formatting, and returns valid and invalid subsets along with structured feedback $\mathcal{F}$:  
\begin{equation}
(D_v, D_{\neg v}, \mathcal{F}) = \mathcal{C}(D_{\text{init}}).
\end{equation}

The invalid subset $D_{\neg v}$ is then re-annotated by the ReasonAgent conditioned on the feedback:
\begin{equation}
D_r \leftarrow \mathcal{R}(D_{\neg v}; \mathcal{F}),
\end{equation}
and the CriticAgent re-validates the regenerated annotations:
\begin{equation}
(D_v', D_{\neg v}', \mathcal{F}') \leftarrow \mathcal{C}(D_r).
\end{equation}

This feedback-regeneration loop continues until $D_{\neg v}$ is empty or the maximum number of iterations is reached, with updates at each iteration:
\begin{equation}
D_v \leftarrow D_v \cup D_v', \quad D_{\neg v} \leftarrow D_{\neg v}', \quad \mathcal{F} \leftarrow \mathcal{F}'.
\end{equation}

The CriticAgent also detects underrepresented regions in the global distribution of intents, attributes and labels:
\begin{equation}
G \leftarrow \mathcal{C}_{\text{dist}}(D_v),
\end{equation}
which are later supplied to the GenAgent to address long-tail data scarcity. The final base dataset after Reason–Critic iterations is $D_b \leftarrow D_v$.

\subsubsection{GenAgent}
For each query $q_i$ in the underrepresented regions $G$, the GenAgent generates supplementary titles to cover insufficient relevance categories (relevant, partially relevant, irrelevant) and attributes. Specifically, for each query $q_i$, it produces titles for underrepresented aspects:
\begin{equation}
T_{\text{gen}} = \mathcal{G}(q_i).
\end{equation}

These synthetic titles are subsequently annotated by the ReasonAgent and validated by the CriticAgent and added to the supplementary dataset $D_s$. This process diversifies coverage, mitigates semantic fragmentation, and ensures long-tail representation.

\subsubsection{Agent Collaboration}
The three agents operate in a closed loop: the ReasonAgent provides reason–label annotations, the CriticAgent validates and balances the dataset, and the GenAgent augments long-tail cases. Iterative refinement continues until convergence. Finally, the CriticAgent integrates all validated samples into the curated dataset $D^*$:
\begin{equation}
D^* \leftarrow \mathcal{C}_{\text{final}}(D_b \cup D_s),
\end{equation}
producing a fully automated, high-quality, diverse, and socially aligned dataset for training.

\subsection{Training Process}
With the curated dataset $D^*$, ASARL performs a three-stage training process that aligns models with social-aware relevance requirements while ensuring deployability.  

\subsubsection{Social Context Training (SCT)}
In the first stage, Supervised Fine-Tuning builds a foundation using $D^*$. Two complementary prompt formats are employed:  
\begin{itemize}
    \item \textbf{Reason+Label Prompt}: requires the model to generate an explanatory reasoning trace $r$ followed by the final relevance label $l$, which encourages explicit modeling of social intent–attribute alignment.
    \item \textbf{Label-Only Prompt}: requires the model to directly output the relevance label $l$, providing a concise and efficient supervision signal. 
\end{itemize}

Let $\pi_\theta$ denote the model parameterized by $\theta$.  
Both prompts are optimized with the standard autoregressive objective:
\begin{equation}
\begin{aligned}
\mathcal{L}_{\mathrm{SCT}} = & \; -\frac{1}{N}\sum_{i=1}^{N} \sum_{t=1}^{T_i} 
\log \pi_\theta\!\big((r_i,l_i)_t \mid q_i, t_i, (r_i,l_i)_{<t}\big) \\
  & - \frac{1}{N} \sum_{i=1}^{N} \log \pi_\theta \left( l_i \mid q_i, t_i\right),
\end{aligned}
\end{equation}
where $(r_i,l_i)_t$ denotes the $t$-th token in the reasoning–label sequence of the $i$-th training sample and $N$ is the number of training samples.  

Reasoning supervision enhances social intent–attribute interpretability, while label-only supervision enforces precision.

\subsubsection{Preference-Guided Optimization (PGO)}
After SCT, which endows the model with socially aware and interpretable reasoning, we further refine it via Preference-Guided Optimization (PGO) to align model outputs with user interaction signals.

Let $D^* = {(q_i, t_i, r_i, l_i)}$ be the curated dataset. We first identify mispredicted samples:
\begin{equation}
\mathcal{M} = {(q_i, t_i, r_i, l_i) \in D^* \mid (r_i, l_i) \neq (\hat{r}_i, \hat{l}_i)},
\end{equation}
where $(\hat{r}_i, \hat{l}_i)$ denotes the model prediction. Labels are verified using user interaction signals (e.g., CTR, join rate) and considered valid if the metrics satisfy predefined thresholds.

For each verified sample, we construct a preference pair $(r_i^+, l_i^+)$, $(r_i^-, l_i^-)$ for optimization. The positive sequence $(r_i^+, l_i^+)$ is generated via beam search under the model $\pi_\theta$, selecting the highest-probability candidate consistent with the verified label $l_i$:
\begin{equation}
(r_i^+, l_i^+) = \arg\max_{(r,l) \in \mathcal{B}(q_i, t_i), l=l_i} \pi_\theta(r,l \mid q_i, t_i),
\end{equation}
where $\mathcal{B}(q_i, t_i)$ is the beam search candidate set. The negative sequence $(r_i^-, l_i^-)$ is taken as the model’s original prediction $(\hat{r}_i, \hat{l}_i)$. This construction preserves label alignment while allowing the reasoning sequence to reflect the model distribution, enabling the DPO loss to effectively align outputs with user preferences.

The model is then optimized via the DPO objective:
\begin{equation}
\begin{aligned}
\mathcal{L}_{\text{PGO}} = 
- \mathbb{E}_{(q,t,(r^+,l^+),(r^-,l^-))} \Big[ 
\log \sigma \Big( 
\beta \log \frac{\pi_\theta(r^+,l^+|q,t)}{\pi_{\text{ref}}(r^+,l^+|q,t)} \\
- \beta \log \frac{\pi_\theta(r^-,l^-|q,t)}{\pi_{\text{ref}}(r^-,l^-|q,t)} 
\Big) 
\Big],
\end{aligned}
\end{equation}
where $\pi_\theta$ is the model being optimized and $\pi_{\text{ref}}$ is the SCT-trained reference model.  

This design ensures that positive samples remain consistent with ground-truth labels yet realizable by the model (via beam search), while negative samples reflect its actual mispredictions, thus enabling DPO loss to effectively align reasoning and labeling with real user preferences.

\subsubsection{Social Distillation (SD)}
Finally, under deployment scenarios with strict latency or resource constraints, we optionally distill the PGO-enhanced teacher model into a lightweight student (BERT). 
Let $l_i$ denote the teacher's predicted relevance label for query–title pair $(q_i, t_i)$, and $p_S(l_i \mid q_i, t_i)$ the student's predicted label distribution.  

Knowledge distillation is performed by minimizing the KL divergence between teacher and student label distributions:
\begin{equation}
\mathcal{L}_{\text{SD}} = \frac{1}{N} \sum_{i=1}^{N} 
\text{KL} \big( s(l_i \mid q_i, t_i) \parallel p_S(l_i \mid q_i, t_i) \big),
\end{equation}
where $s(l_i \mid q_i, t_i)$ is either the softmax of teacher logits or the one-hot label for hard-label distillation.

This selective distillation transfers alignment and preference-aware knowledge into compact models while providing flexible trade-offs between social-aware accuracy and deployment efficiency.


\section{Experiments}
This section reports offline and online evaluations conducted to assess the effectiveness of our proposed framework. We first describe the experimental setup (datasets, metrics, baselines, and implementation details), then present the offline and online results, and finally provide qualitative case analyses.

\subsection{Experimental Setup}
\subsubsection{Datasets}
The datasets were extracted from QQ search logs, totaling approximately 1.1M query–title pairs after being annotated using the agent system. Both the training and test splits are drawn from the same overall distribution. To provide broad domain coverage, queries were sampled uniformly from predefined categories (e.g., education, gaming, health). For each query, 10 candidate titles were sampled from three tiers: top-ranked cases, token-matched cases, and randomly selected cases. The dataset covers two QQ search scenarios—\emph{Group Search} and \emph{Channel Search}—with examples from both. The test set has been manually reviewed to ensure annotation accuracy. Summary statistics are presented in \Cref{tab:datasets}.
\begin{table}[h]
  \centering
  \caption{Statistics of the datasets.}
  \label{tab:datasets}
  \begin{tabular}{cccc}
    \toprule
    \textbf{Scene} & \textbf{Train} & \textbf{Validation} & \textbf{Test} \\
    \midrule
    Group   &602281  &60395  &59182  \\
    Channel &512039  &50824  &51363  \\
    \bottomrule
  \end{tabular}
\end{table}
\subsubsection{Metrics}
We evaluate our models using both offline and online metrics to provide a comprehensive assessment of performance. For offline evaluation, we adopt standard relevance metrics. Macro F1 measures the harmonic mean of precision and recall across all classes, treating each class equally. NDCG assesses the quality of the predicted ranking of titles relative to the ground-truth relevance labels. Accuracy computes the proportion of correctly predicted relevance labels over the entire test set. 

For online evaluation, we adopt three metrics. Click-Through Rate (CTR) measures the proportion of clicks per query impression. Join Rate (JR) captures how often users join a group or channel after a search interaction, reflecting deeper engagement. To compare our model against the baseline, we report GSB, which balances the number of improved versus degraded cases; a higher score indicates that the model yields more consistent gains. Together, these offline and online metrics provide a holistic view of both the predictive accuracy and the real-world user impact of the models.

\subsubsection{Baselines}
We compare our framework with a range of strong baselines, including classical pre-trained models and large language models. To ensure fairness, each baseline is evaluated under both unfine-tuned and fine-tuned settings.  

\begin{itemize}
    \item \textbf{BERT}\cite{devlin2019bert}: In the unfine-tuned setting, BERT is used to compute query–title similarity directly. In the fine-tuned setting, the model is trained with a classification objective (\texttt{[CLS]} token) to predict the relevance score.  
    \item \textbf{RoBERTa}\cite{liu2019roberta}: Similar to BERT, the unfine-tuned model is applied for similarity matching, while the fine-tuned variant is trained with a classification loss to predict relevance.  
    \item \textbf{LLM Base}: We adopt the Qwen3-0.6B\cite{yang2025qwen3} model as the representative LLM baseline. In the unfine-tuned setting, the model directly outputs a relevance label in a zero-shot manner. In the fine-tuned setting, it is further trained on our dataset to generate the correct label.  
    \item \textbf{ASARL (SCT)}: Our proposed ASARL, initialized from Qwen3-0.6B and Qwen3-8B, both of which were fine-tuned with Social Context Training on the curated training data.
    \item \textbf{ASARL (SCT+PGO)}: An enhanced version of ASARL that further applies Preference Guided Optimization to mitigate bias and better align model outputs with human preferences.  
\end{itemize}

\subsubsection{Implementation Detail}
The multi-agent system is equipped with tools that enable socially aware data curation: a social relevance criteria for intent–attribute alignment, a predefined sub-table of social attribute categories (Table~\ref{tab:attributes}), access to the curated dataset $D^*$ and memory buffers for read/write operations, and statistical utilities to monitor label and attributes distributions and detect underrepresented regions for long-tail augmentation. All agents—ReasonAgent, CriticAgent, and GenAgent—are implemented using Qwen3-235B as the underlying model. These capabilities allow the agents to collaboratively generate, validate, and enrich the dataset while maintaining social relevance and distributional balance.

\begin{table}[h]
\centering
\caption{Categorization of 10 structured social attributes on the QQ search platform.}
\label{tab:attributes}
\begin{tabular}{|c|c|c|c|c|c|}
\hline
Film   & Lifestyle       & Industry & Education  & Gaming \\ \hline
Location & Gender       & Age & Knowledge &Otaku    \\ \hline
\end{tabular}
\end{table}

All training is conducted on 4 NVIDIA A100 GPUs with a batch size of 16. We adopt AdamW as the optimizer with a learning rate of $1\times10^{-5}$ and apply a warm-up ratio of 0.05. The input sequence length is capped at 1024 tokens, and models are trained for 4 epochs. To mitigate overfitting, we monitor validation performance using a 10\% split of the training data and retain the best-performing checkpoint. For the PGO stage, the training is limited to 1 epochs to further reduce overfitting risks. During inference, we employ vLLM for efficient deployment, with the temperature fixed to 0 to ensure deterministic predictions.

\subsection{Offline Evaluation}
To comprehensively assess the effectiveness of our Autonomous Social-Aware Relevance Learning (ASARL), we conduct offline experiments to quantify model performance on relevance prediction.

\subsubsection{Overall Performance.} 
\Cref{tab:overall_performance} compares all baseline models and our ASARL variants under two settings: without fine-tuning (w/o) and with fine-tuning (w/). Without fine-tuning, BERT, RoBERTa and LLMBase achieve relatively low MacroF1, NDCG, and Accuracy, indicating that pre-trained models alone are insufficient to capture the complex relevance patterns in social search, which depend on contextual cues, domain-specific terminology, and latent user intent.  

Fine-tuning on the curated social search dataset improves all models. Our proposed ASARL models exhibit the most significant gains. $\text{ASARL}^{\text{SCT}}$, trained with supervised fine-tuning on carefully curated social search data, substantially outperforms other fine-tuned baselines, demonstrating the importance of high-quality data and constructed COT supervision. Further integrating Preference Guided Optimization in $\text{ASARL}^{\text{SCT+PGO}}$ leads to additional improvement, suggesting that aligning the model with user interaction preferences provides a more nuanced understanding of relevance beyond what is captured by labeled data alone. Moreover, the $\text{ASARL}_{8\text{B}}$ variants consistently outperform their $\text{ASARL}_{0.6\text{B}}$ counterparts, showing that larger models benefit more from fine-tuning and preference optimization.

\begin{table}[t]
  \centering
  \caption{Overall performance comparison of baseline models and ASARL variants under two settings: without fine-tuning (w/o) and with fine-tuning (w/).}
  \label{tab:overall_performance} 
  \begin{tabularx}{\linewidth}{clccc} 
    \toprule
    \textbf{Finetune}& \textbf{Method} & \textbf{MacroF1$\uparrow$} &\textbf{NDCG@4$\uparrow$} & \textbf{ACC$\uparrow$} \\
    \midrule
    \multirow{3}{*}{w/o} 
      & BERT &60.73  &67.78  &63.03  \\
      & RoBERTa &60.54  &68.15  &63.89  \\
      & LLMBase & 59.36 & 66.43 & 64.80 \\
    \midrule
    \multirow{5}{*}{w/} 
      & BERT &66.23  &75.13  &67.92  \\
      & RoBERTa &66.19  &75.88  &68.30  \\
      & LLMBase & 68.35 & 74.92 & 72.56 \\
      & $\text{ASARL}_{0.6\text{B}}^{\text{SCT}}$ & 75.59 & 76.84 & 77.34 \\
      & $\text{ASARL}_{0.6\text{B}}^{\text{SCT+PGO}}$ & 75.65 & 77.01 & 77.90 \\
      & $\text{ASARL}_{8\text{B}}^{\text{SCT}}$ & 82.72 & 77.25 & 83.85 \\
      & $\text{ASARL}_{8\text{B}}^{\text{SCT+PGO}}$ &\textbf{83.66}  &\textbf{77.93}  &\textbf{84.52}  \\
    \bottomrule
  \end{tabularx}
\end{table}

\subsubsection{Effectiveness of Each Agent} 
\Cref{tab:Agent Performance} presents the performance of the multi-agent pipeline under incremental ablations: ReasonAgent alone, ReasonAgent with CriticAgent, and the full pipeline including GenAgent. Using ReasonAgent only yields the lowest MacroF1, NDCG, and Accuracy, indicating that while ReasonAgent can generate initial relevance annotations, errors in reasoning consistency, logic, and output formatting limit its effectiveness.  

Adding CriticAgent improves all three metrics. By providing feedback to ReasonAgent, CriticAgent enables iterative correction of inconsistencies and logical errors, with an observed correction rate of approximately 21\%. This feedback loop results in higher-quality annotations and increased precision, as reflected in the improved MacroF1 and Accuracy.  

Incorporating GenAgent further enhances all metrics, achieving the highest MacroF1, NDCG, and Accuracy in the full three-agent setup (Reason+Critic+Gen). This improvement reflects GenAgent's role in generating additional high-quality examples, particularly for long-tail queries, thereby increasing coverage and diversity.  

\begin{table}[h]
  \caption{Impact of Reason, Critic, and Gen agents on overall effectiveness.}
  \label{tab:Agent Performance}
  \renewcommand{\arraystretch}{1.1} 
  \begin{tabular}{ 
    >{\centering\arraybackslash}p{2.5cm} 
    >{\centering\arraybackslash}p{1.6cm} 
    >{\centering\arraybackslash}p{1.6cm} 
    >{\centering\arraybackslash}p{1.3cm} }
    \toprule
    \textbf{Agents}& \textbf{MacroF1$\uparrow$} &\textbf{NDCG@4$\uparrow$} & \textbf{ACC$\uparrow$} \\
    \midrule
    Reason &70.96&76.04&73.39\\
    Reason+Critic &71.88&76.23&73.90\\
    Reason+Critic+Gen &\textbf{73.55}&\textbf{76.41}&\textbf{75.75}\\
  \bottomrule
\end{tabular}
\renewcommand{\arraystretch}{1}
\end{table}

\subsubsection{Effective of constructed COT Tuning.} We further investigate the effect of constructed Chain-of-Thought (CoT) tuning on relevance prediction (\Cref{tab:COT Performance}). Compared to direct prediction (Label), a generic CoT yields only marginal improvement in MacroF1 and Accuracy  while slightly decreasing NDCG, indicating that unguided step-by-step reasoning does not reliably capture social search subtleties. In contrast, our Social-Aware CoT—whose reasoning steps are explicitly designed to reflect social signals such as group culture, implicit intent, and attribute alignment—delivers substantially larger gains. Combining Social-Aware CoT with label supervision produces the best results, which shows that domain-specific reasoning and supervised labels are complementary: the former injects social priors and interpretable cues, while the latter grounds the model with explicit relevance judgments.

\begin{table}[h]
  \caption{Effectiveness of Social-Aware versus generic CoT reasoning in relevance prediction.}
  \label{tab:COT Performance}
  \begin{tabularx}{\linewidth}{cccc}
    \toprule
    \textbf{Method} & \textbf{MacroF1$\uparrow$} &\textbf{NDCG@4$\uparrow$} & \textbf{ACC$\uparrow$} \\
    \midrule
    Label  &68.35&74.92&72.56\\
    Generic CoT &69.02&74.76&72.89\\
    Social-Aware CoT &73.55&76.41&75.75\\
    Social-Aware CoT + Label &\textbf{75.59}&\textbf{76.84}&\textbf{77.34}\\
  \bottomrule
  \end{tabularx}
\end{table}

\subsubsection{Effectiveness of Knowledge Distillation.} 
To evaluate the impact of our multi-stage training pipeline, we compare the online model (RoBERTa) with the version distilled after SCT and PGO. As shown in Table~\ref{tab:knowledge_distill}, the distilled model consistently outperforms the baseline across Macro-F1, NDCG, and Accuracy. This demonstrates that our pipeline effectively transfers richer supervision and preference signals into the student model. While distillation reduces the model size and complexity, it results in some performance trade-offs, but still retains sufficient effectiveness for real-world deployment.

\begin{table}[h]
  \centering
  \caption{Comparison between the RoBERTa-online baseline and the distilled model.}
  \label{tab:knowledge_distill}
  \renewcommand{\arraystretch}{1.1} 
  \begin{tabular}{ 
    >{\centering\arraybackslash}p{2.5cm} 
    >{\centering\arraybackslash}p{1.6cm} 
    >{\centering\arraybackslash}p{1.6cm} 
    >{\centering\arraybackslash}p{1.3cm} }
    \toprule
    \textbf{Models} & \textbf{MacroF1$\uparrow$} & \textbf{NDCG@4$\uparrow$} & \textbf{ACC$\uparrow$} \\
    \midrule
     RoBERTa-online & 65.01 & 75.70 & 67.45 \\
     RoBERTa-distilled & \textbf{72.05} & \textbf{76.41} & \textbf{73.92} \\
    \bottomrule
  \end{tabular}
  \renewcommand{\arraystretch}{1}
\end{table}

\begin{figure*}[t]
  \centering
  \includegraphics[width=\linewidth]{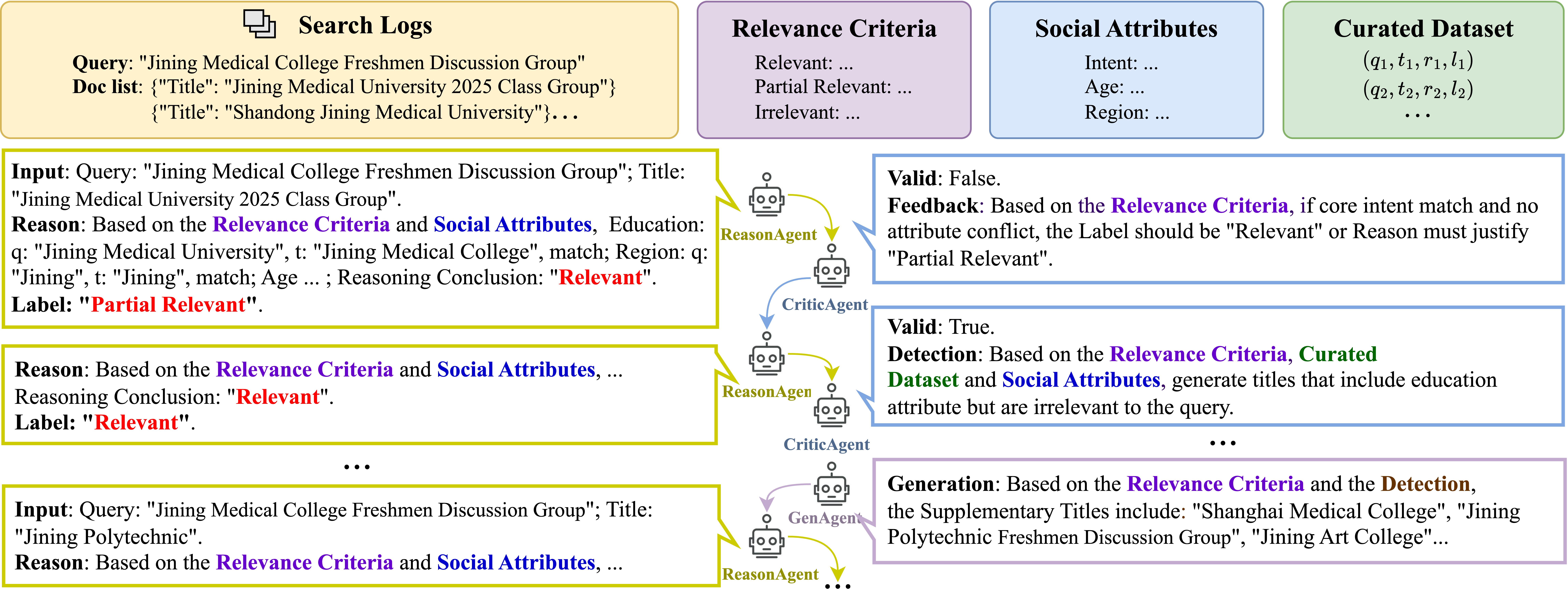}
  \caption{Illustration of the collaborative workflow among ReasonAgent, CriticAgent, and GenAgent. The ReasonAgent performs initial reasoning and labeling, the CriticAgent detects inconsistencies and enforces alignment, and the GenAgent generates supplementary examples to enrich dataset diversity.}
  \label{fig:agents}
\end{figure*}

\subsection{Online Evaluation}
Online deployment are conducted to measure real-world user engagement and business metrics.
\subsubsection{Online Deployment.}
To assess our framework, we deployed it on the QQ A/B test platform. We assigned 20\% traffic to the experimental group (using ASARL) and another 20\% to the baseline group (using the previous RoBERTa-based model). For fair comparison, we continuously monitored performance and set a minimum 7-day test period to mitigate traffic fluctuation impacts.

\subsubsection{Online Performance.}
The ASARL framework was deployed in both the \emph{channel} and \emph{group} search scenarios, yielding consistent improvements across key metrics. In the channel search scenario, CTR, JR, and GSB increased by 2.69\%, 2.59\%, and 11.66\%, respectively. Similarly, in the group search scenario, the corresponding gains were 1.36\%, 1.06\%, and 16.66\%. These results highlight that ASARL not only enhances user engagement (CTR), but also drives stronger interaction (JR) and broader social growth (GSB), thereby improving both user satisfaction and platform-level outcomes. The deployment reached a DAU of 12 million, validating its scalability.

\subsection{Case studies}


\subsubsection{Collaboration between agents.} Agent collaboration is essential for ensuring both annotation quality and dataset diversity. As illustrated in Figure~\ref{fig:agents}, a typical scenario arises when the ReasonAgent produces inconsistent outputs between the reasoning process and the assigned label. For instance, given the pair (query: "Jining Medical College Freshmen Discussion Group", title: "Jining Medical University 2025 Class Group"), the reasoning indicates a relevance of "Relevant", whereas the final label is "Partial Relevant". The CriticAgent identifies this inconsistency and enforces alignment, thereby maintaining logical closure. 

After examining the data distribution, the CriticAgent further instructs the GenAgent to generate supplementary examples under specific attributes (education) and relevance level (irrelevant). In this case, the GenAgent produces titles such as "Jining Polytechnic Freshmen Discussion Group" and "Shanghai Medical College", which capture fine-grained social characteristic distinctions. This collaborative workflow underscores the complementary roles of ReasonAgent, CriticAgent, and GenAgent in maintaining social and logical consistency while enriching the dataset.

\subsubsection{Training Impact on Relevance Judgment} We present a representative case in \Cref{tab:case_SCT_dpo_attributes} that shows how SCT and PGO change model’s social-aware understanding. The query is "post-00s dating and friend-expansion group" and the title is "friend-making and partner-seeking group".
Before training, the base model misidentifies the query intent yet still considers it matched with the title, and it also incorrectly infers that the title contains the "post-00s" age marker. Together, these errors lead the model to assign label "Relevant".
After SCT and PGO, the trained model recognizes the semantic similarity in intent and correctly detects that the "post-00s" age attribute appears only in the query and not in the title. As a result, the model downgrades the judgment to label "Partial Relevant".
This case illustrates that SCT and PGO reduce over-reliance on lexical similarity and improve social-level discrimination, yielding more accurate and fine-grained relevance decisions.

\definecolor{Match}{RGB}{198,239,206}      
\definecolor{Mismatch}{RGB}{255,199,206}   
\definecolor{Partial}{RGB}{255,235,156}    

\newcommand{\cmark}{\textcolor{green!70!black}{\ding{51}}} 
\newcommand{\xmark}{\textcolor{red}{\ding{55}}} 

\begin{table}[h]
\centering
\caption{Effect of SCT and PGO training in social context understanding and reasoning. Query: "post-00s dating and friend-expansion group"; Title: "friend-making and partner-seeking group".}
\label{tab:case_SCT_dpo_attributes}
\renewcommand{\arraystretch}{1.2}
\begin{tabular}{|>{\centering\arraybackslash}m{1.2cm}|>{\raggedright\arraybackslash}m{4.9cm}|>{\centering\arraybackslash}m{1.3cm}|}
\hline
\textbf{Model} & \textbf{Reasoning} & \textbf{Label} \\ \hline
LLMBase &
\textbf{Intent}: ...\colorbox{Match}{match} \cmark \newline 
$q$: "post-00s" / $t$: "friend-making" \xmark \newline
\textbf{Age}: ...\colorbox{Match}{match} \xmark \newline 
$q$: "post-00s" / t: "post-00s" \xmark &
"Relevant" \\ \hline
ASARL &
\textbf{Intent}: ...\colorbox{Match}{match} \cmark \newline 
$q$: "friend-expansion", "dating" / $t$: "making friends", "partner-seeking" \cmark \newline
\textbf{Age}: ...\colorbox{Mismatch}{mismatch} \cmark \newline 
$q$: "post-00s" / $t$: "—"  \cmark &
"Partial Relevant" \\ \hline
\end{tabular}
\end{table}


\section{Conclusion}
In this work, we present \textbf{ASARL}, an autonomous social-aware relevance learning framework that combines multi-agent data curation with staged model training. By leveraging ReasonAgent, CriticAgent, and GenAgent, ASARL generates socially grounded, interpretable data and augments long-tail queries. Meanwhile, its three-stage training pipeline, Social Context Training (SCT), Preference-Guided Optimization (PGO), and Social Distillation (SD), enables effective social context learning from curated data and behavioral signals. Extensive experiments on the QQ platform demonstrate improvements in relevance prediction, user engagement, and annotation efficiency, validating ASARL as a scalable and interpretable solution for social search.

\bibliographystyle{ACM-Reference-Format}
\bibliography{sample-base}


\end{document}